\documentclass[structabstract]{aa}  
\usepackage{graphicx}
\usepackage{psfig}
\usepackage{natbib}
\usepackage[english]{babel}  
\usepackage{txfonts}
\usepackage{url}

\usepackage{txfonts}
\begin{document}
   \title{X-raying hot plasma in solar active regions with the SphinX spectrometer}

   \author{M. Miceli
          \inst{1,2}
          \and
          F. Reale\inst{1,2}
          \and
          S. Gburek\inst{3}
          \and
          S. Terzo\inst{2}
          \and
          M. Barbera\inst{1,2}          
          \and
          A. Collura\inst{2}
          \and
	  J. Sylwester\inst{3}
	  \and
	  M. Kowalinski\inst{3}
	  \and
	  P. Podgorski\inst{3}
	  \and
	  M. Gryciuk\inst{3}
	  }

   \institute{Dipartimento di Fisica, Universit\`a di Palermo,
              Piazza del Parlamento 1, 90134 Palermo\\
              \email{miceli@astropa.unipa.it}
         \and
             INAF-Osservatorio Astronomico di Palermo, Piazza del Parlamento 1, 90134 Palermo\\ 
         \and
             Space Research Centre, Polish Academy of Sciences, 51-622, Kopernika 11, Wroc\l aw, Poland}
   \date{Received September 15, 1996; accepted March 16, 1997}

% \abstract{}{}{}{}{} 
% 5 {} token are mandatory
 
  \abstract
  % context heading (optional)
  % {} leave it empty if necessary  
   {}
  % aims heading (mandatory)
   {The detection of very hot plasma in the quiescent corona is important for diagnosing heating mechanisms. The presence and the amount of such hot plasma is currently debated. The SphinX instrument on-board CORONAS-PHOTON mission is sensitive to X-ray emission well above 1 keV and provides the opportunity to detect the hot plasma component. } 
  % methods heading (mandatory)
   {We analyzed the X-ray spectra of the solar corona collected by the SphinX spectrometer in May 2009 (when two active regions were present). We modelled the spectrum extracted from the whole Sun over a time window of 17 days in the $1.34-7$ keV energy band by adopting the latest release of the APED database.}
  % results heading (mandatory)
   {The SphinX broadband spectrum cannot be modelled by a single isothermal component of optically thin plasma and two components are necessary. In particular, the high statistics and the accurate calibration of the spectrometer allowed us to detect a very hot component at $\sim7$ million K with an emission measure of $\sim2.7\times10^{44}$ cm$^{-3}$.  The X-ray emission from the hot plasma dominates the solar X-ray spectrum above 4 keV. We checked that this hot component is invariably present both at high and low emission regimes, i.e. even excluding resolvable microflares. We also present and discuss a possible non-thermal origin (compatible with a weak contribution from thick-target bremsstrahlung) for this hard emission component.}
  % conclusions heading (optional), leave it empty if necessary 
   {Our results support the nanoflare scenario and might confirm that a minor flaring activity is ever-present in the quiescent corona, as also inferred for the coronae of other stars.}


   \maketitle
%
%________________________________________________________________

\section{Introduction}
\label{sec:intro}

The solar corona is X-ray bright. The bulk of the emission is thermal, with a spectrum made of a multitude of emission lines, e.g. of iron, overlapping a bremsstrahlung continuum, characteristic of an optically thin plasma at 1-3 MK. Most of the emission comes from plasma confined in closed magnetic tubes anchored in the photosphere, the coronal loops. Hotter plasma, to temperatures well above 10 MK, is typically detected during very impulsive flare events, occurring in localized coronal regions.

Recently, the high sensitivity of new imaging and spectral instrumentation has allowed to refine the investigation of the thermal structure of the corona. One question is the presence of minor very hot component at temperature above 6-8 MK even outside of macroscopic flare events. This question is linked to the more basic question of the presence of flaring activity down to very small scales, to constitute a major steady component of the heating extending to the entire solar corona.
Although very faint, such component has been detected in the recent past in active regions both with filter ratios or differential emission measure (DEM) reconstruction from imaging broad-band instruments (Hinode/XRT, e.~g., \citealt{Reale2009a,Reale2009b,Mctiernan2009a,Schmelz2009b}) and with spectroscopic line analysis (single line or DEM reconstruction, e.~g., \citealt{Ko2009a,Patsourakos2009a,Shestov2010a,Sylwester2010a}). More solid confirmation has been found very recently \citep{Testa012a}. The weight of this very hot component has been estimated to be a few percent at most of the major component around 3 MK in an active region \citep{Reale2009a}. However, there are reasons to believe that the amount of very hot plasma may vary from active region to active region and it may be smaller in less prominent active regions \citep[][e.g.]{Testa2011a}.

Yet, another question becomes the weight of this hot component over long timescales and over the global coronal budget.
Some hints come from Sun-as-a-Star analysis with Yohkoh data \citep{Argiroffi2008a} but with large uncertainties in the hotter part. More tight constraints are then needed in the hot tail.

The Solar PHotometer IN X-rays (SphinX) (\citealt{skk08}, \citealt{gsk11}), part of the THESIS package on board the CORONAS-PHOTON solar mission, is appropriate to investigate this issue: it was a broad-band spectrometer (1-15 keV), with moderate spectral resolution but sensitive up to high energies, enough to detect possible small hot components. \citet{skg12} (hereafter SKG12) analyzed coronal spectra measured with SphinX at the last solar minimum and obtained that they are well-described by a thermal spectrum at about 1.7-1.9 MK, with significant flux only in the 1-3 keV range.

Here we analyze the full-disk unresolved corona during a low solar activity period but in the presence of active regions, with the aim to search for the detection of a possible minor hot component. In this analysis, we take advantage of the possibility to integrate over several observation days. This allows us to achieve a very high signal-to-noise ratio and to access to higher energies ($>3$ keV), where the steep coronal spectrum yields a very low count rate, but which is crucial to be sensitive to a hot emission measure component. 

Although we include active regions in the analysis, we address the non-flaring corona and therefore in the data selection we exclude any major flare event. We will discuss about the role of minor, still resolved, flaring components (microflares). 

Going to such high energies one may wonder about the possible role of non-thermal emission. High-energy spectrum components are for instance typically detected with RHESSI during flares \citep[][e.g.]{Veronig2006a,Krucker2008a}. In the case of lower activity, we may hope in principle to detect non-thermal emission in the SphinX spectral range. So we pay attention also to this issue.

The paper is organized as follows: in Sect. \ref{DA} we describe the data and the data analysis procedure; in Sect. \ref{Results} we show the results of the spectral analysis (and of the count-rate resolved spectral analysis). Our conclusions are discussed in Sect. \ref{Discussion}.

\section{Data Analysis}
\label{DA}

The spectra presented in this paper were collected by the D1 detector of the SphinX spectrometer. The detector energy range is $1.2-14.9$ keV and the spectral resolution is $\sim460$ eV. Further details on the instrumental characteristics can be found in \citet{gsk11} and in SKG12.

\begin{figure}[htb!]
 \centerline{\hbox{     
     \psfig{figure=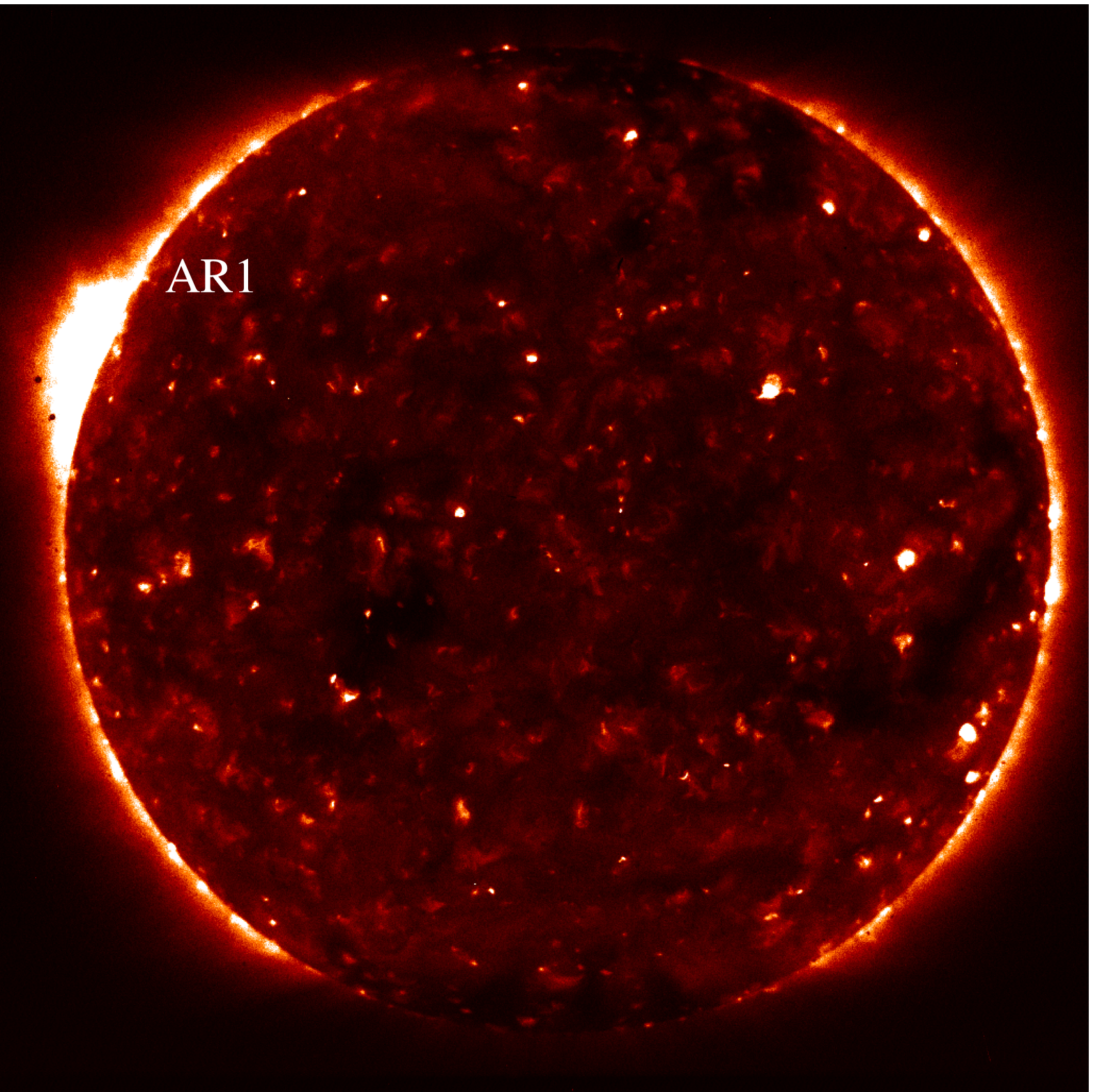,width=7cm}}}
      \centerline{\hbox{     
     \psfig{figure=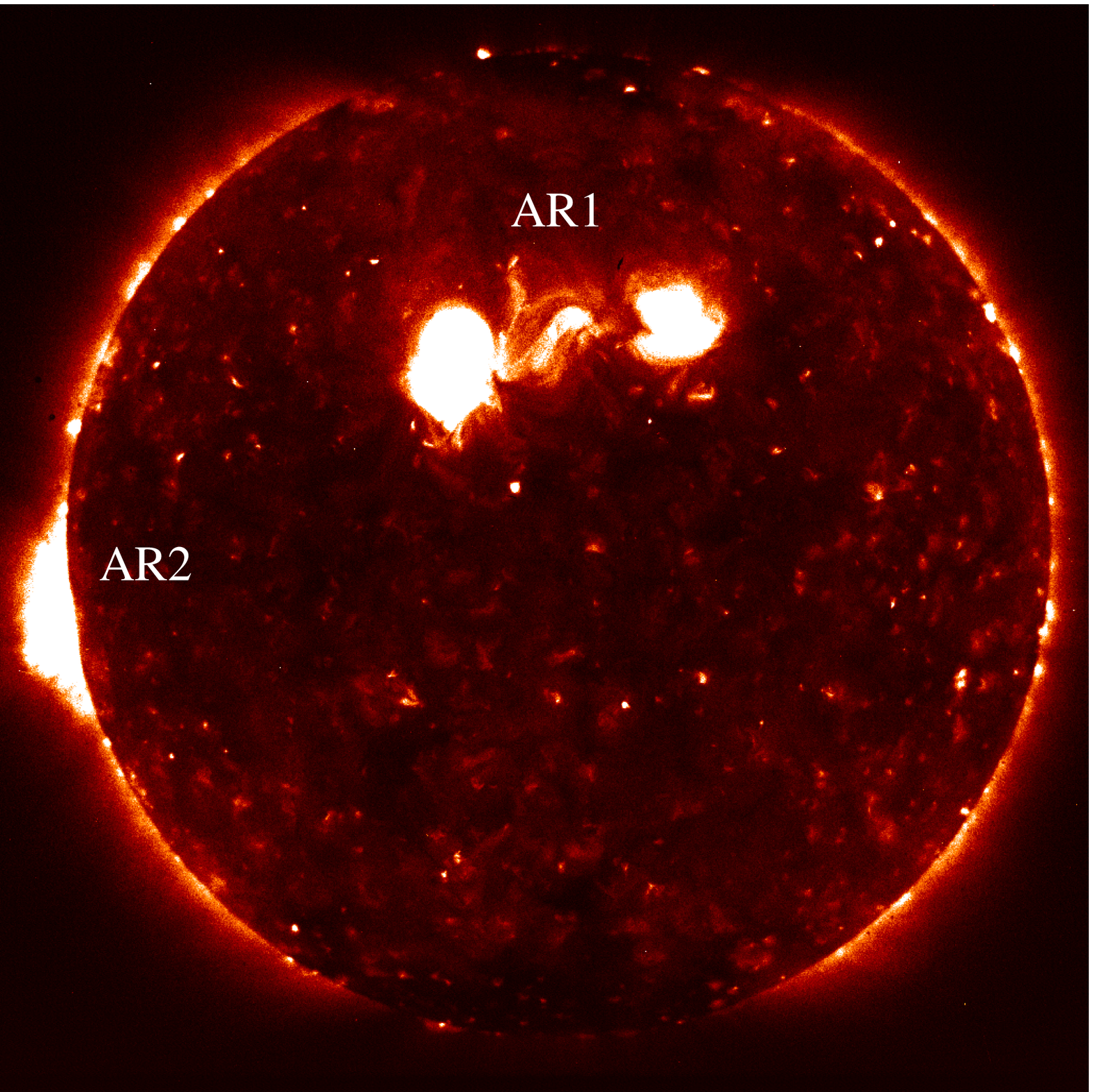,width=7cm}}}  
        \centerline{\hbox{     
     \psfig{figure=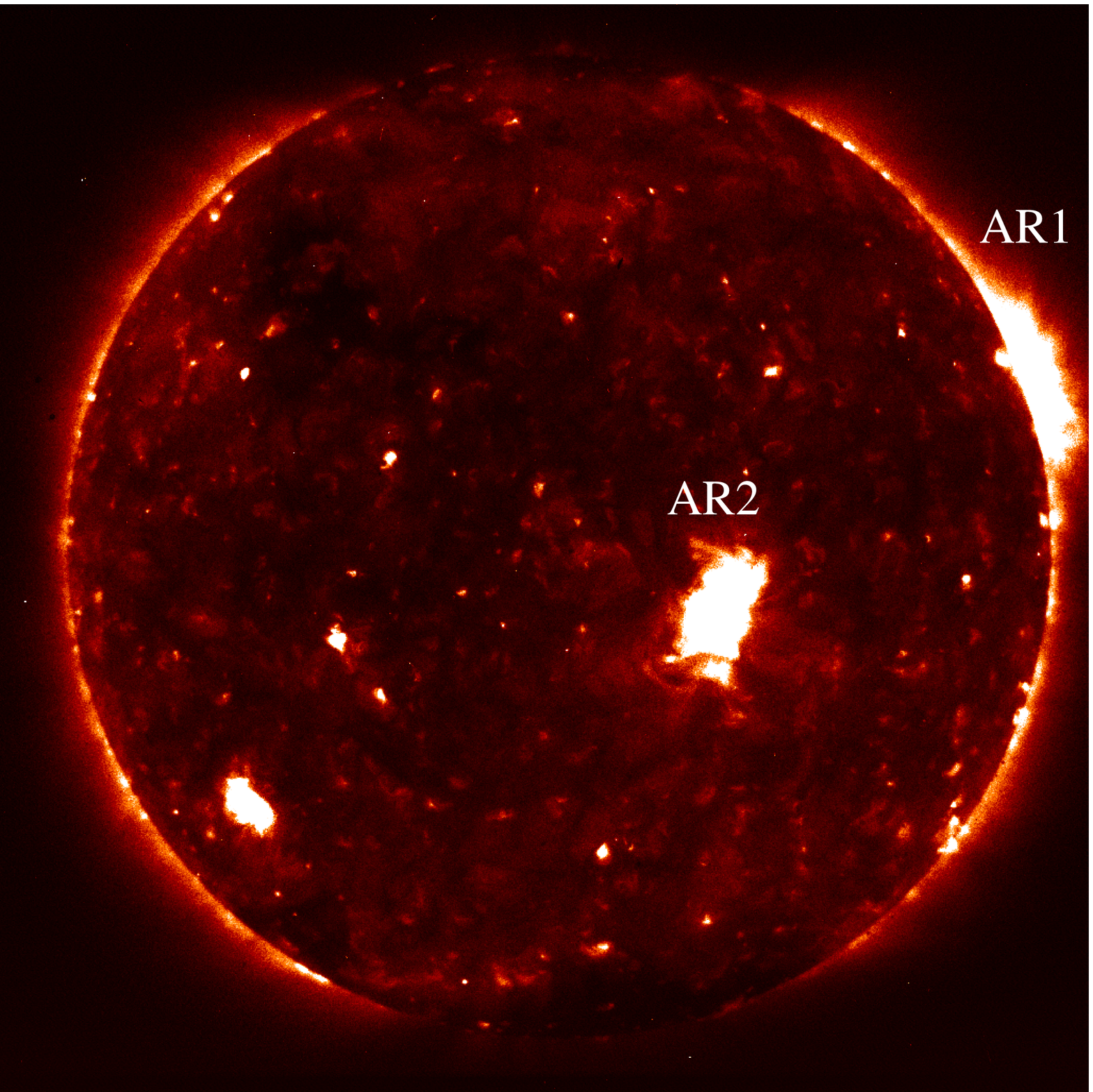,width=7cm}   
 }}
\caption{$Hinode/XRT$ synoptic images of the solar corona (Ti$-$poly filter) in 2009 May 7 (\emph{upper panel}), 2009 May 15 (\emph{central panel}), and 2009 May 23 (\emph{lower panel}). The position of active regions AR1 and AR2 is marked.}
\label{fig:XRT}
\end{figure}

We considered all the observations between 2009 May 7 and 2009 May 24 (the FITS event files are available at the SphinX data catalogue website: \url{http://156.17.94.1/sphinx_l1_catalogue/SphinX_cat_main.html}). The choice of this time window was dictated by the relatively  high solar X-ray flux associated with the presence of active regions and by the lack of significant flaring events. The active region AR11017 (hereafter AR1) was visible between May 7 and 22, while a second, dimmer, active region (hereafter AR2) was visible from May 15 to the end of our time window. Figure \ref{fig:XRT} shows the $Hinode/XRT$ synoptic images of the Sun (Ti$-$poly filter) in May 7, May 15, and May 23 together with the corresponding positions of AR1 and AR2.

To reduce a possible contamination from solar flares, we carefully inspected the light-curves of all the observations and excluded observations SPHINX\_090509\_192730\_254950 and SPHINX\_090514\_062135\_064952 to remove a B1.0 and a A5.9 flare, respectively. 

For each event file we extracted the corresponding spectrum by applying a very conservative screening criterium to remove particle-related events, spurious mesurements, and non-GTI events. Namely, we selected FLAG=0 events only (see the SphinX web site for a detailed description of individual flags). We then summed the single spectra to obtain a total spectrum with more than $3.9\times10^7$ counts and 57 ks of filtered exposure time. 

We adopted the response matrix SPHINX\_RSP\_256\_nom\_D1.fts and performed spectral analysis by using XSPEC V12.7 (\citealt{arn96}). We performed gain fits within Xspec to account for possible weak variations of the response gain as an effect of fluctuations in the detector temperature during the 17 days spanned by our observations. This procedure shifts the energies $E$ on which the response matrix is defined according to the formula $E'=(E/g_1) - g_2$, where $(g_1,g_2)$ are free parameters in the fit. For all the models described below we found $g_1\sim1.025$, $g_2\sim-0.03$, in agreement  with the expectations of small variations in the response matrix.

Spectral analysis was performed in the $1.34-7$ keV energy band (i. e. instrument channels $24-121$) to minimize calibration issues ($1.2-1.34$ keV) and contaminations from the electronic noise ($7-14.9$ keV).
The total spectrum was modelled by adopting the APEC spectral code (optically thin coronal plasma in collisional ionization equilibrium, \citealt{sbl01}) based on the 2.0 release of the AtomDB database (see \url{http://www.atomdb.org}). Because of the relatively limited SphinX spectral resolution it was not possible to clearly identify line emission features in the spectra and to perform accurate diagnostics of the abundances. Therefore, we set the abundance table in the spectral model to that reported by \citet{fel92} for the solar upper atmosphere. We added a systematic 5\% error term in our spectral fittings to account for the estimated uncertainties in the calibration of the spectrometer. 
% (elements not listed therewith were set to the photospheric values according to \citealt{ag89}).

\section{Results}
\label{Results}

\begin{figure}[htb!]
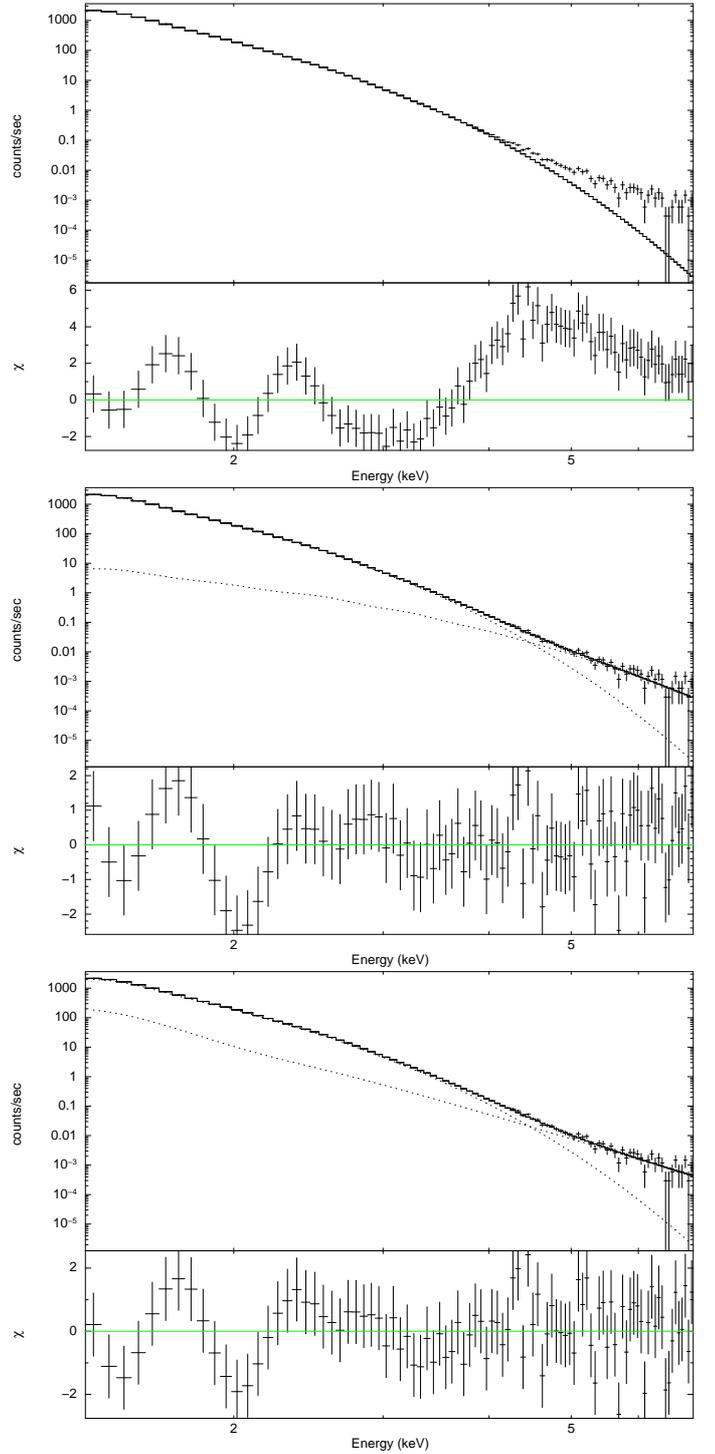

 \centerline{\hbox{     
     \psfig{figure=1T.ps,angle=-90,width=\columnwidth}}}
      \centerline{\hbox{     
     \psfig{figure=2T.ps,angle=-90,width=\columnwidth}}}  
      \centerline{\hbox{     
     \psfig{figure=1Tpow.ps,angle=-90,width=\columnwidth}}}  
     \caption{\emph{Upper panel:} SphinX spectrum of the solar corona collected between 2009 May 7 and 2009 May 24 together with its best-fit model consisting of one isothermal component and residuals. \emph{Middle panel:} same as upper panel with a two-component thermal model. The contribution of each component is shown. \emph{Lower panel:} same as middle panel with the 1T-Pow model.}
\label{fig:spec}
\end{figure}

We first verified that the total SphinX broad-band spectrum cannot be properly described by a single isothermal component. A fit of this simple model to the spectrum provides an unacceptable reduced $\chi^2=7.06$ (with 93 d.~o.~f.) and significant residuals are visible in the high-energy tail of the spectrum, as shown in the upper panel of Fig. \ref{fig:spec}. 

The SphinX spectrum can be well fitted by adding a second thermal component (see middle panel of Fig. \ref{fig:spec}). The model with two components provides a reduced $\chi^2=1.07$ (with 91 d.~o.~f.) and a null hypothesis probability $>30\%$. The $1-8$ $\AA$ X-ray flux is $1.4\times 10^{-5}$ erg$/$s$/$cm$^2$ and the $1-15$ keV luminosity is $4.7\times10^{23}$ erg$/$s. Table \ref{tab:res} shows the best-fit parameters obtained with one and two thermal components.
We did not obtain significant improvements in the quality of the fits by adding a further thermal component. In particular, according to the F-test, the probability that the improvement in the fit is not significant is more than three times higher than our threshold at $0.1\%$ and the error bars in the best-fit parameters grow larger. We therefore conclude that the broadband modeling of the SphinX spectrum requires two thermal components, a warm component at $\sim2.7\times10^6$ K and a hot one $\sim7\times10^6$ K  (hereafter 2T model).

\begin{center}
\begin{table}[htb!]
\begin{center}
\caption{Best-fit parameters obtained by modelling the SphinX spectrum with one (1T) and two (2T) isothermal components of optically thin plasma, and with one thermal component plus a power-law (1T-Pow). All errors are at the 90\% confidence level.}
\begin{tabular}{lccc} 
\hline\hline
 {\bf Total Spectrum}            &        1T              &        2T                  &  1T-Pow \\ \hline
 $T_1$ ($10^6$ K)                & $2.83^{+0.02}_{-0.01}$ &        $2.73\pm0.01$       & $2.73\pm0.01$  \\ 
 $EM_1$ ($10^{46}$ cm$^{-3}$)    &      $53.6\pm+1.3$     &          $63\pm2$          & $63\pm2$   \\
 $T_2$ (K)                       &              -         &     $6.6^{+0.4}_{-0.2}$    &      -      \\
 $EM_2$ ($10^{46}$ cm$^{-3}$)  &              -         & $0.027^{+0.009}_{-0.001}$   &        -    \\
 $\Gamma$                       &              -         &           -                 &  $9.0\pm0.3$  \\
$N_{pow}^{*}$                     &               -           &           -                 &  $7^{+4}_{-2}\times10^4$  \\
  $\chi^2$ (d~.o.~f.)            &     $657.0$ $(93)$     &         $97.1$ $(91)$      &  $94.0$ $(91)$        \\
\hline\hline
\multicolumn{4}{l}{\footnotesize{* Photons$/$s$/$keV$/$cm$^2$ at 1 keV.}} \\
\label{tab:res}
\end{tabular}
\end{center}
\end{table}
\end{center}

The high-energy tail of the SphinX spectrum can also be fitted by a non-thermal component. A model with a (warm) thermal component plus a power-law (1T-Pow model, see lower panel of Fig. \ref{fig:spec}) provides a $\chi^2=94.0$ (with 91 d. o. f.) that is even lower than that obtained with the 2T model. In this case, the values of $T_1$ and $EM_1$ are consistent with those obtained with the 2T model, while for the non-thermal component we obtain a quite steep photon index $\Gamma$, as shown in Table \ref{tab:res}.
%These values are in agreement with those expected (VEDI REF FABIO)la
Therefore, with the available data it is not possible to exclude that the hard X-ray emitting component is associated with non-thermal processes and$/$or is a combination of thermal and non-thermal emission.

SKG12 have analyzed SphinX spectra extracted from the lowest-activity periods of the 2009 solar minimum and characterized by the lack of  major active regions. They found that in time intervals of exceptionally low X-ray flux, SphinX spectra can be modelled by a single thermal component at $1.7-1.9\times 10^6$ K with emission measure ranging between $4\times10^{47}$ cm$^{-3}$ and $1.1\times10^{48}$ cm$^{-3}$.
We note that in the spectra studied by SKG12 the $1-15$ keV X-ray luminosity is a factor $6-10$ lower than that in our spectrum. The differences in X-ray luminostity, plasma temperature, and spectral model can be related to the presence of the active regions AR1 and AR2 in our observations. We examined the $Hinode/XRT$ synoptic images of the Sun in the thin$-$Be filter, whose bandwidth is approximatively $0.9-2.5$ keV (three observations available in our time window: May 16, 19, and 20) and found that the contribution (in terms of DN detected by XRT) of AR1 and AR2 is only $\sim6-9\%$ of the total. Therefore, the contribution of the active regions to the total flux dominates at high energy only.

In principle it is possible to ascertain the contribution of AR1 and AR2 by assuming that the spectrum of the underlying corona is described by the model adopted by SKG12. Under this assumption one can add a third, cooler, component to the 2T model (by forcing in the fitting process that its temperature, $T_0$, and its emission measure, $EM_0$, are in the ranges derived by SKG12). Unfortunately this approach is not feasible since, as explained above, the addition of a new component makes the measures of the best-fit parameters less precise. Nevertheless, this 3T model yelds $T_0\sim1.9\times10^6$ K and $EM_0\sim 1\times10^{48}$ cm$^{-3}$, while $T_1$, $T_2$, $EM_1$, and $EM_2$ are all consistent with those reported in Table \ref{tab:res} at less than one sigma (but with much larger uncertainties).

\begin{figure}[htb!]
 \centerline{\hbox{     
     \psfig{figure=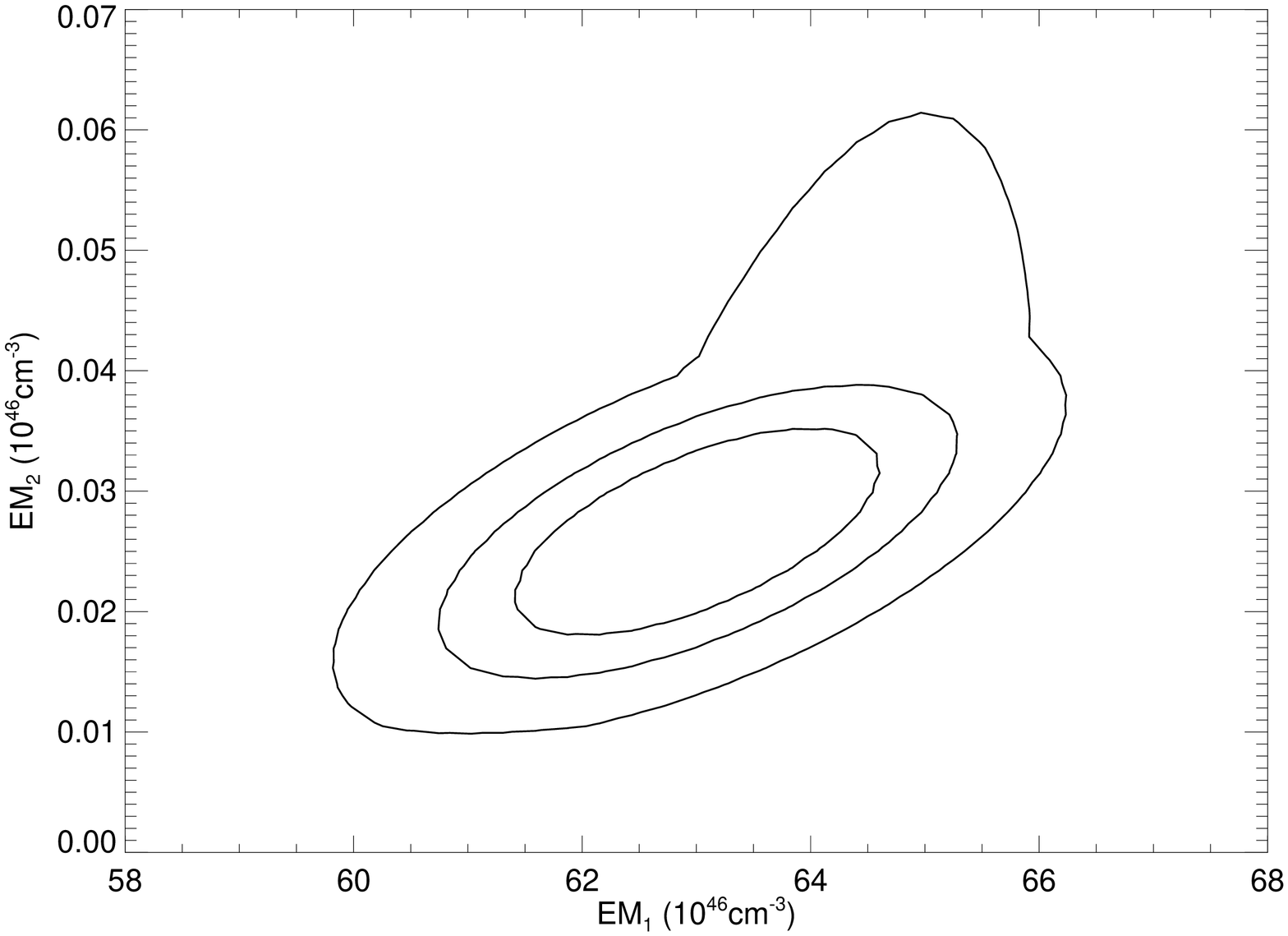,angle=90,width=\columnwidth}}}
      \centerline{\hbox{     
     \psfig{figure=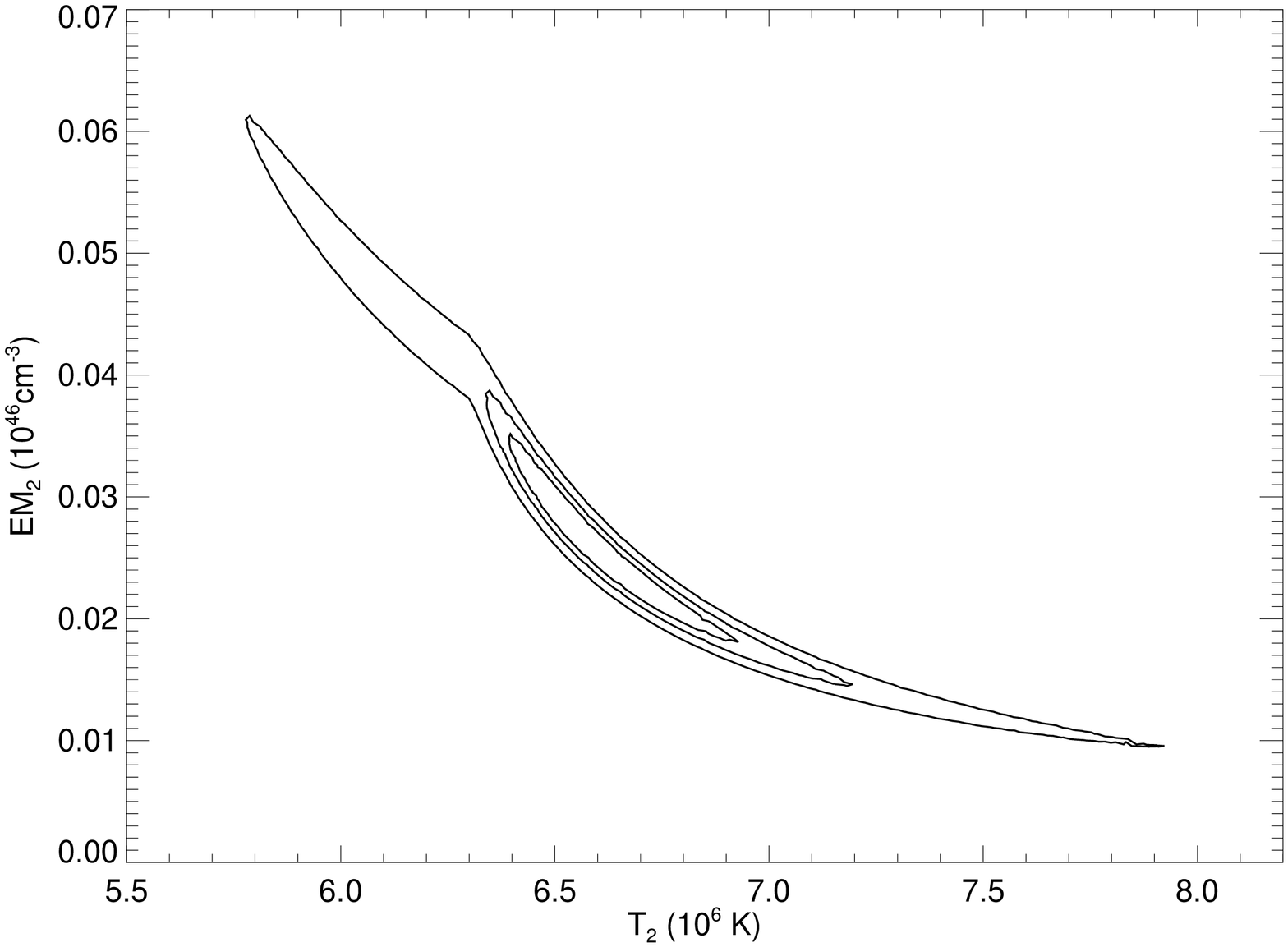,angle=90,width=\columnwidth}}}  
\caption{\emph{Upper panel:} $68\%$, $90\%$, and $99\%$ confidence contour levels of the emssion measure of the hot component, $EM_2$, versus the emission measure of the warm component $EM_1$ obtained with the 2T model \emph{Lower panel:} Same as upper panel for $EM_2$ and $T_2$.}
\label{fig:step}
\end{figure}

Upper panel of Fig. \ref{fig:step} shows the $68\%$, $90\%$, and $99\%$ confidence contour levels of the emssion measure of the warm component, $EM_1$, versus the emission measure of the hot component $EM_2$. The contours indicate that the ratio $EM_1/EM_2$ is almost constant at $\sim2500$. The best-fit value of $T_2$ is entangled with $EM_2$, as shown by the $68\%$, $90\%$, and $99\%$ confidence contour levels in lower panel of Fig. \ref{fig:step}.

\subsection{Count-rate resolved spectral analysis}

\begin{figure}[htb!]
 \centerline{\hbox{     
     \psfig{figure=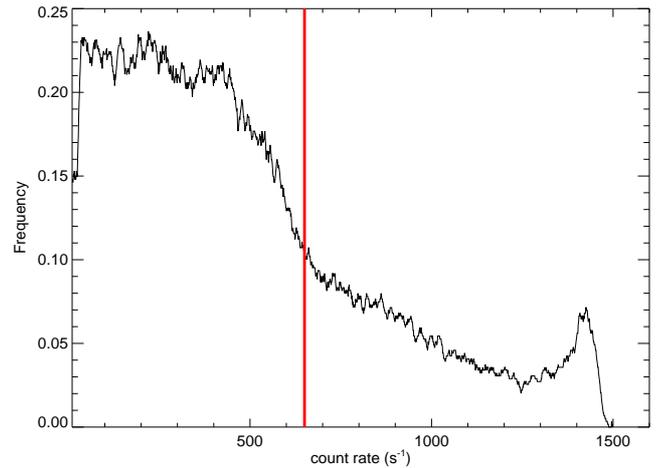,angle=90,width=\columnwidth}}}
\caption{Distribution of the X-ray count-rate observed by SphinX between 2009 May 7 and 2009 May 24. The red vertical line indicates the threshold chosen to extract the low count-rate spectrum and the high count-rate spectrum (see text).}
\label{fig:histo}
\end{figure}

\begin{figure}[htb!]
  \centerline{\hbox{	 
      \psfig{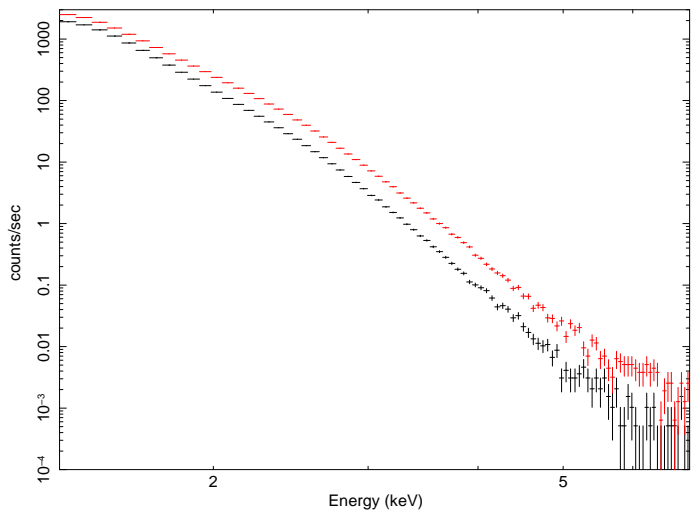}}}
 \caption{SphinX spectrum of all the events detected at low count-rate ($<650$ s$^{-1}$, black crosses) and at high count-rate ($>650$ s$^{-1}$ red crosses) between 2009 May 7 and 2009 May 24.}
\label{fig:lrhr}
\end{figure}

Figure \ref{fig:histo} shows the distribution of the X-ray count-rate observed by SphinX in our 17 days time window. The average count-rate is of a few hundreds counts s$^{-1}$, but a peak at $\sim1400$ counts s$^{-1}$ is also visible.
The peak at high count-rate may be associated with microflares not directly visible through a simple inspection of the light curve. These microflares may also be responsible for the hot component detected in the solar spectrum. To investigate this issue we performed a count-rate resolved spectral analysis by extracting the spectrum of all the events detected at low count-rate ($<650$ s$^{-1}$, hereafter $LR$ spectrum) and the spectrum of all the events detected at high count-rate ($>650$ s$^{-1}$, $HR$ spectrum). Figure \ref{fig:lrhr} shows the $LR$ and $HR$ spectra.

We found a clear relationship between count-rate and spectral hardness, with the $HR$ spectrum being significantly harder than the $LR$ spectrum. Namely, the soft ($1.34-3$ keV) to hard ($3-7$ keV) flux ratio is $\sim 1000$ in the $LR$ spectrum and $\sim500$ in the $HR$ spectrum.
This result supports a possible link between high count-rate and microflare activity. 
Nevertheless, though the $LR$ spectrum is softer than the $HR$ spectrum, the former cannot be modelled by a single thermal component (reduced $\chi^2=3.57$ with 93 d.~o.~f.) and a proper fit still requires an additional hot component (reduced $\chi^2=0.98$ with 91 d.~o.~f.), or a power-law (reduced $\chi^2=1.01$ with 91 d.~o.~f.).
This clearly proves that the hot component cannot be only an effect of microflares, since it is present even in regimes of low count-rate, where any possible contamination from microflares is removed.

Table \ref{tab:reslh} shows the best-fit parameters obtained by fitting the $LR$ and $HR$ spectra with one and two thermal components, and with one thermal component plus a power-law. Because of the lower statistics, error bars are somehow larger than those obtained for the total spectrum and we still found an entanglement between the best-fit values of $EM_2$ and $T_2$ (similar to that shown in Fig. \ref{fig:step}). Table \ref{tab:reslh} shows that the best-fit temperatures obtained for the $HR$ spectrum are higher than those obtained for the $LR$ spectrum (and from those obtained for the total spectrum, see Table \ref{tab:res}). 
%Also, in terms of emission measure, the weight of the hot HR component is larger than that of the hot LR component. 
This is coherent with the $HR$ spectrum showing harder X-ray emission.

\begin{center}
\begin{table}[htb!]
\begin{center}
\caption{Best-fit parameters obtained by modelling the $LR$ and $HR$ spectrua (see text) with one (1T) and two (2T) isothermal components of optically thin plasma, and with one thermal component plus a power-law (1T-Pow). All errors are at the 90\% confidence level.}
\begin{tabular}{lccc} 
\hline\hline
 {\bf LR Spectrum}             &        1T              &        2T                  &   1T-Pow   \\ \hline
 $T_1$ ($10^6$ K)                & $2.71\pm0.01$       &        $2.65\pm0.01$        &   $2.65\pm0.01$          \\ 
 $EM_1$ ($10^{46}$ cm$^{-3}$)  &      $54.6\pm1.2$     &          $61\pm2$           &      $61\pm2$    \\
 $T_2$ (K)                       &              -         &     $6.4^{+0.6}_{-1.0}$    &       -        \\
 $EM_2$ ($10^{46}$ cm$^{-3}$)  &              -         & $0.018^{+0.03}_{-0.009}$  &          -           \\ 
  $\Gamma$                       &              -         &           -                 &  $8.9^{+0.4}_{-0.5}$  \\
$N_{pow}^{*}$      &  -      &           -                 &  $4^{+3}_{-2}\times10^4$  \\ \hline
  $\chi^2$ (d~.o.~f.)            &     $332.5$ $(93)$     &         $89.6$ $(91)$      &  $91.9$ $(91)$  \\
\hline\hline
 {\bf HR Spectrum}             &        1T              &        2T                  &    1T-Pow       \\ \hline
 $T_1$ ($10^6$ K)                & $3.21^{+0.04}_{-0.01}$ &        $3.17\pm0.01$       &  $3.17\pm0.01$    \\ 
 $EM_1^*$ ($10^{46}$ cm$^{-3}$)  &  $40.6^{+0.4}_{-0.6}$  &     $43.0^{+0.7}_{-0.6}$  & $43.0^{+0.7}_{-0.6}$ \\
 $T_2$ (K)                       &              -         &     $8.0^{+0.4}_{-1.2}$    &     -   \\
 $EM_2^*$ ($10^{46}$ cm$^{-3}$)  &              -         & $0.019^{+0.03}_{-0.006}$  &      -     \\ 
  $\Gamma$                       &              -         &           -                 &  $8.2\pm0.3$  \\
$N_{pow}^{*}$                  &            -      &           -                 &  $5\pm2\times10^4$  \\\hline
  $\chi^2$ (d~.o.~f.)            &     $488.3$ $(93)$     &         $102.2$ $(91)$      &  $102.7$  $(91)$\\
\hline \hline
\multicolumn{4}{l}{\footnotesize{* Photons$/$s$/$keV$/$cm$^2$ at 1 keV.}} \\
\label{tab:reslh}
\end{tabular}
\end{center}
\end{table}
\end{center}

%Finally we investigated the possible non-thermal origin for the hard X-ray emission in the $LR$ and $HR$ spectra. The 1T-Pow model provides $\chi^2=91.9$ and $\chi^2=102.7$ (with 91 d.~o.~f.) for the $LR$ and $HR$ spectra, respectively, and the best-fit values of $T_1$ and $EM_1$ are consistent  with those obtained with the 2T model for both spectra. We obtained $\Gamma=8.9^{+0.4}_{-0.5}$ and $N_{pow}=4^{+3}_{-2}\times10^4$ cm$^{-2}$ keV$^{-1}$ s$^{-1}$ for the $LR$ spectrum, and $\Gamma=8.2\pm0.3$ and $N_{pow}=5\pm2\times10^4$ cm$^{-2}$ keV$^{-1}$ s$^{-1}$ for the $HR$ spectrum.

\section{Discussion and conclusions}
\label{Discussion}

We have illustrated the analysis of full-disk X-ray data collected with SphinX spectrometer over a two weeks period when one or two active regions were ever present on the solar disk.
Although SphinX has no spatial resolution, its broad band and sensitivity up to high X-ray energies ($>10$ keV)  make it suited to investigate the interesting question of the presence and role of high temperature plasma components in the solar corona.

Although the solar X-ray spectrum is quite steep, integrating over two weeks allows us to reach a very high signal-to-noise ratio well above 3 keV. This makes a real difference from previous studies. We analyze such a broad-band spectrum, with one and two-thermal component models and we obtain that the spectrum is not well-described by a single thermal component and a second hotter component improves the fitting at high significance level. In practice, in addition to the large warm component previously found in the very quiet sun,  that we confirm ($T \approx 2.7 $ MK), our analysis detects a minor hot component ($T \approx 7$ MK) at extremely high significance. The X-ray emission from the hot plasma dominates the solar X-ray spectrum above 4 keV and it is clearly associated with the presence of the active regions, that were absent in the analysis of SKG12. The temperature of this hot component is compatible with that found in other studies of active regions \citep{Reale2009a,Testa2011a}. The results of 
present 
investigations  are also in an agreement with the multitemperature analysis (DEM reconstruction) based on the 15 line fluxes observed by RESIK spectrometer in the range 3.5 and 6 $\AA$ \citep{Sylwester2010a}: the examination of non flaring periods (activity classes from A9 to B5) at the beginning of 2003 indicated that temperatures of hot DEM component are between 6 MK and 9 MK and its emission measure is 3 orders of magnitude smaller than that of the cooler component.

Our fittings show that the emission measure of the hotter component is less than 0.1\% of the other cooler one. As mentioned in Sect.\ref{sec:intro}, in other active regions there was evidence of a hot component of the order of a few percent of the main warm one. Although this evidence was subject to important uncertainties and needed quantitative confirmations, there are reasons that may help to understand even this amount of discrepancy. The active regions studied previously were quite intense ones. Less intense active regions did not show this amount of hot plasma, and might be more compatible with the amount that we find \citep{Testa2011a}. Similarly, here we are close to the solar minimum, and these active regions are not very intense. Also, here we are analyzing full-disk observations, including the quiet Sun emission. This emission may contribute significantly to the detected warm emission and much less to the hot one, making the ratio hot$/$warm decrease. Finally, we integrate over two weeks and 
therefore we average spatially but also on a very long time baseline, that may include periods when the hot component is relatively high and periods when it is not. In other words, we provide the time-average of the hot component whereas the other measurements were snapshot. This time-averaged value might be interesting when evaluating the importance of heating mechanisms that involve the systematic presence of hot plasma, as nanoflare theory (\citealt{Parker1988a, Cargill1994a, Klimchuk2006a}).

In any case, although at such small amount, the hot plasma detected by our analysis is at extremely high significance and shows that the data collected with SphinX are highly sensitive to this component.

One important question about the hot component is whether it might be due to minor, still resolvable, microflares or to continuous, widespread and very small nanoflares. 
The Soft X-ray Telescope (SXT) on Yohkoh found microflares to occur in active regions in the $0.25-4$ keV band \citep{shi95}. Also RHESSI observes microflares exclusively in active regions, although with different rates and in the hard X-rays.
Microflares observed systematically with RHESSI typically show similar spectral features as those reported in Fig. \ref{fig:spec} and Fig \ref{fig:lrhr} although in a harder energy band (\citealt{hck08,hhb11}). The different energy bands correspond to different fitting parameters, as expected from experimental bias \citep{hhb11}. We can equally attempt a comparison if we either consider microflares included in or excluded from our data.
The microflares analyzed by \citet{lsk84} show nonthermal emission with $\Gamma = 4-6$ above 30 keV, i.e. at considerably higher energies with respect to the SphinX energy band.
The electron spectrum can be assumed to exist above some ``cutoff'' energy EC , which means the accelerated electron population appears abruptly at energies where the thermal spectrum is negligible. For microflares, the observed break tends down to very low energies where there are multiple emission lines in the thermal spectra. This makes it very difficult to determine EC accurately and to determine the energy in a microflare's nonthermal electrons, and this holds also for our analysis. A systematic analysis of a large sample of microflares observed with RHESSI has shown an average power law index $\Gamma = 6.9$, with most events in the range $4 < \Gamma < 10$ (\citealt{hck08}). Our nonthermal fitting falls in the right tail of this distribution, i.e. in the low-energy tail.
The microflare spectra collected by RHESSI are often so steep that they are difficult to distinguish from the thermal emission (\citealt{hck08}), as it occurs in our analysis. Thermal fitting yield temperatures above 10 MK with an average of about 13 MK (\citealt{hck08}). These temperatures are significantly higher than those obtained for the hot component in our analysis. Part of the dicrepancy might be explained with the different sensitivity of the instruments, due to the different spectral band, that would make them give different weights to a broad emission measure distribution (\citealt{por00}, \citealt{hhb11}). The average emission measure obtained from thermal fitting of RHESSI microflares is about $2-3 \times 10^{46}$ cm$^{-3}$, much larger than the emission measure of the hot components that we obtain. We should consider that our value is obtained as an average value over a long time baseline. Since the microflaring is not continuous, for a sound comparison we should scale the values to include the 
microflare duty cycle. If we assume a 10\% duty cycle (\citealt{hhb11}), our value still remains quite lower than expected if microflares are included.

Nevertheless, we have tried to disentangle a possible contribution from microflare activity with an independent approach, namely we analyzed separately the spectra obtained from the low and high count-rate periods. We obtained a very robust answer: the hot component is confirmed at high significance in both regimes, with the only difference of noticeably different temperature values and emission measures, but that do not affect the general result. Therefore, we find this hot component at any level of emission intensity.

Finally, we have also examined the possibility that this hot component could be instead due to non-thermal emission. Although on the basis of other previous studies the expectation that the emission is thermal is solid, the data do not allow us to exclude that the excess in the hard energy tail is due to non-thermal emission. Fitting with a non-thermal component provides sound results, i.e. a power law with quite a steep spectral index, still compatible with a weak contribution from thick-target bremsstrahlung \citep{Veronig2006a}.
Nevertheless, the 1T-Pow framework would imply the lack of hot X-ray emitting plasma from active regions at odd with previous studies.
Discriminating between thermal and non-thermal emission requires at the same time high sensitivity and good spectral resolution, to resolve possible high-energy lines such as the Fe XXV 6.7 keV line. Even in that case we probably could not exclude a simultaneous presence of thermal and non-thermal spectral components.
We point out that if part of the hard X-ray emission that we detected  has a non-thermal origin, the $EM$ values for the hot component reported in Table \ref{tab:res} and in Table \ref{tab:reslh} might be considered as upper limits.
In both cases of thermal and non-thermal emission this invariably brings support to the presence of steady impulsive heating in active regions. 

In conclusion, this analysis of SphinX data confirms the presence of a minor yet steady very hot component in coronal active regions, therefore supporting an important role of impulsive processes in bright plasma heating. Future higher resolution and high sensitivity X-ray spectroscopy might shed more light on the details and nature of this hot emission.

\begin{acknowledgements}
We thank the anonimous referee for their useful comments and suggestions. M.~M, F.~R, and S.~T acknowledge support from Italian Ministero dell'Universit\`a e Ricerca and 
from Agenzia Spaziale Italiana (ASI)/INAF agreement, contract I/023/09/0. S.~G. and J.~S. acknowledge financial support from the Polish Ministry of Education and Science Grant 2011/01/B/ST9/05861. The research leading to these results has received funding from the European Commission's Seventh Framework Programme (FP7/2007-2013) under the grant agreement eHeroes (project n° 284461, www.eheroes.eu). Hinode is a Japanese mission
developed and launched by ISAS/JAXA, with NAOJ as domestic partner and NASA and STFC (UK) as international partners.
It is operated by these agencies in cooperation with ESA and
NSC (Norway).

\end{acknowledgements}

\bibliographystyle{aa}
%\bibliography{refs_cite.bib}

%\begin{thebibliography}{}
%\end{thebibliography}

\end{document}